\documentclass[prb,preprint]{revtex4-1}

\pdfoutput=1

\usepackage{subcaption}

\usepackage{color}
\usepackage{graphicx}
\usepackage{amsmath}
\usepackage{amsfonts}
\usepackage{amssymb}

\usepackage{url}

\usepackage[colorlinks=true
,urlcolor=blue
,anchorcolor=blue
,citecolor=blue
,filecolor=blue
,linkcolor=blue
,menucolor=blue
,linktocpage=true
,pdfproducer=medialab
,pdfa=true
]{hyperref}

\begin{document}

\title{Observation of relativistic corrections to Moseley's law at high atomic number}

\author{Duncan C. Wheeler}
\email{duncanw@alum.mit.edu}
\author{Emma Bingham}
\email{bingham@mit.edu}
\author{Michael Winer}
\author{Janet M. Conrad}
\author{Sean P. Robinson}
\affiliation{Dept.~of Physics, Massachusetts Institute of Technology, Cambridge, MA 02139, USA}

\date{\today}

\begin{abstract}
Transitions between low-lying electron states in atoms of heavy elements lead to electromagnetic radiation with discrete energies between about 0.1~keV and 100~keV (x rays) that are characteristic of the element. 
Moseley's law --- an empirical relation first described by Moseley in 1913 which supported predictions of the then-new Bohr model of atomic energy levels while simultaneously identifying the integer atomic number $Z$ as the measure of nuclear charge --- predicts that the energy of these characteristic x rays scales as $Z^2$. The foundational nature of Moseley's experiment has led to the popularity of Moseley's law measurements in undergraduate advanced laboratory physics courses. We report here observations of deviations from Moseley's law in the characteristic $K_{\alpha}$ 
x-ray emission of 13 elements ranging from $Z=29$ to $Z=92$. 
While following the square-law predictions of the Bohr model fairly well at low $Z$, the deviations become larger with increasing $Z$ (negligible probability of the Bohr model fitting data by a $\chi^2$ test). We find that relativistic models of atomic structure are necessary to fit the full range of atomic numbers observed (probability value of $0.20$ for the relativistic Bohr-Sommerfeld model). As has been argued by previous authors, measurements of the relativistic deviations from Moseley's law are both pedagogically valuable at the advanced laboratory level and accessible with modern but modest apparatus. Here, we show that this pedagogical value can be be extended even further --- to higher $Z$ elements, where the effects are more dramatically observable --- using apparatus which is enhanced relative to more modest versions, but nevertheless still accessible for many teaching laboratories.
\end{abstract}

\maketitle

\section{Introduction}

Experimental tests of Moseley's law\cite{moseley1, moseley2} using x-ray fluorescence spectroscopy appear commonly in advanced undergraduate physics laboratory courses. Such experiments allow students to explore key supporting evidence for Bohr's atomic theory\cite{Bohr} while also introducing modern precision spectroscopy techniques.  However, experiments testing Moseley's law have the potential to teach undergraduates much more.  Recent work by Soltis \textit{et al.}\cite{soltis2017}\ shows that within the precision of modern detectors, Moseley's law becomes inaccurate at high atomic number and requires first-order relativistic corrections.  By performing a Moseley's law experiment, students receive the opportunity to identify limitations in a commonly taught model, providing insight into the nature of experimental physics.  In this paper, we build upon the work of Soltis \textit{et al.}\ by measuring elements with even higher atomic number which deviate even further from Moseley's law.  We show that the first-order relativistic approximation used by Soltis \textit{et al.}\ remains inaccurate for these heavier elements and find that a more exact relativistic Bohr-Sommerfeld model is required, providing more aspects of the physics and modeling for students to explore.

\section{Background}

In 1913, H. G. J. Moseley experimentally measured the wavelengths of characteristic x~rays from a series of elements. Using his data in conjunction with Bohr's recent theory describing the hydrogen atom, Moseley proposed that the energy of the transition scales quadratically with the atomic number $Z$.\cite{moseley1, moseley2, Bohr}  This quadratic relation, called Moseley's law, formed some of the first observational evidence for a quantum theory of atomic structure.

There are a number of ways to produce x rays in nature. They range from fluorescence to synchrotron radiation to extreme blue-shifting of radio waves. In this experiment, we focus on the first of these. X-ray fluorescence typically occurs when an electron is knocked out of a low-lying shell of a heavy element, leaving a hole in the electronic structure. This could be a consequence of bombardment by alpha rays, beta rays, gamma rays, or some more exotic process. The resulting hole is most often filled with an electron from a nearby higher shell, emitting a photon in the process. 

That electron leaves a new hole, which is then filled in much the same way, resulting in a cascade of electrons between the quantized energy levels of the atom, each emitting a photon. The brightest line in this spectrum comes from electrons transitioning between energy levels in the $n=2$ and $n=1$ shells, where $n$ is the usual principal quantum number. These $2\rightarrow1$ transition x rays are denoted $K_\alpha$ in Siegbahn notation. Bohr's model predicts that electrons in shell $n$ have velocity $Z\alpha c/n$, where $Z$ is the atomic number, $\alpha$ is the fine structure constant (approximately $1/137$\cite{pdb}) and $c$ is the speed of light. This implies that the $K_\alpha$ transition has energy 
\begin{equation}
E_{K_{\alpha}}^\textrm{Bohr}=\frac{3}{8}  m_ec^2\alpha^2(Z-1)^2 \label{eq:bohr},
\end{equation} 
where $Z$ has been replaced by $Z-1$ to account for the effective screened nuclear charge experienced by the electron.

However, Bohr's model also predicts that for a reasonably heavy element like gold ($Z=79$), the innermost electrons are moving at more than half the speed of light. Therefore, it is possible that relativistic corrections may come into play. Soltis \textit{et al.}\ included the first-order perturbative relativistic corrections to Bohr's model (that is, the power series expansion of the kinetic energy, spin-orbit coupling, and the Darwin term) and found 
\begin{equation}
E_{K_{\alpha}}^\textrm{(1)}=m_ec^2\left(\frac {3}{8}\alpha^2 (Z-1)^2+\frac{15}{128}\alpha^4 (Z-1)^4\right) \label{eq:soltis}.
\end{equation}

By using a relativistic Bohr-Sommerfeld approximation, we will find model fits to our data that improve upon the perturbation methods used by Soltis \textit{et al}. This can help students understand the intricacies behind combining quantum mechanical theories with relativistic theories.

\section{Relativistic Bohr-Sommerfeld approximation}

\begin{figure}
\includegraphics[width=9cm]{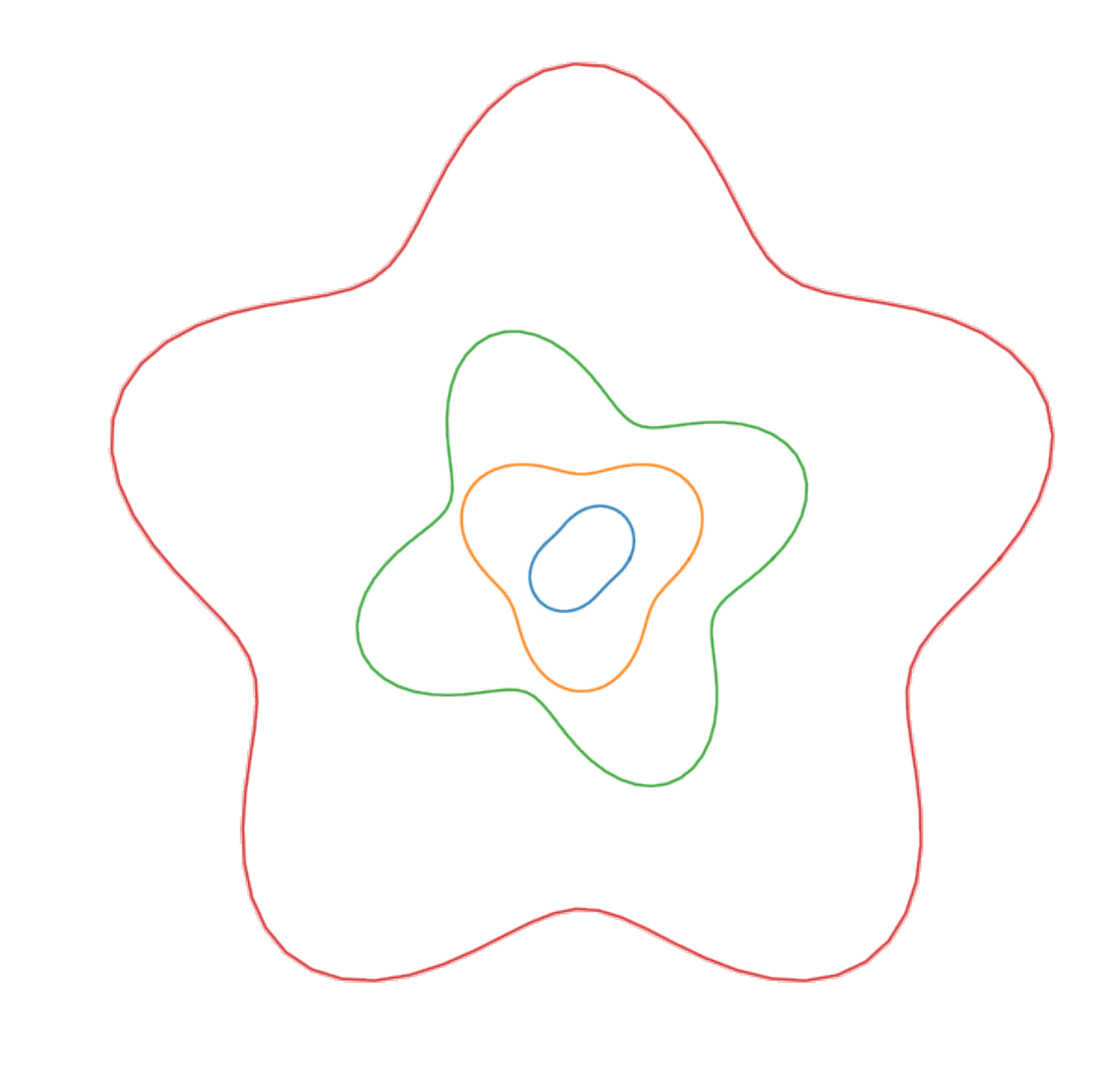}
\caption{The $n=2$, $3$, $4$, and $5$ orbitals in the Bohr-Sommerfeld picture. The sinusoidal radial wiggles do not represent variations in orbital radius, but rather the phase of the wave.}
\label{fig:bohr}
\end{figure}

To better understand these relativistic corrections, we utilize the Bohr-Sommerfeld approximation. The Bohr-Sommerfeld quantization condition is a semiclassical rule that says in any closed orbit in a quantum system $\int _{\textrm{orbit}} pdx=2\pi n\hbar$ for an integer $n$, where $p$ is the system momentum, $x$ the coordinate, and $\hbar=h/2\pi$ is the reduced Planck's constant. If momentum can be thought of as the derivative of a quantum wavefunction's phase, then this condition says that closed orbits are standing waves where the phase is the same at the beginning and end, as shown in Fig.~\ref{fig:bohr}. For circular orbits, the condition requires that angular momentum be $L=n\hbar$. This approximation can be combined with classical mechanics to derive Bohr's atomic theory.

In the relativistic case, we still have $L=n\hbar$. However, $L$ is now $\gamma m_e rv$ instead of $m_e rv$, where $m_e$ is the mass of the electron, $v$ is the velocity, $r$ is the radius of the orbit, and the shorthand $\gamma\equiv1/\sqrt{1-v^2/c^2}$ indicates the Lorentz gamma factor. The result\cite{kraft1974} is 
\begin{equation}
E_n=m_e c^2\sqrt{1-\left(\frac{\alpha(Z-1)}{n}\right)^2} ,
\label{eq:1}
\end{equation}
where $E_n$ is the total energy of the system's $n^\mathrm{th}$ energy eigenstate, including the electron mass-energy. The x-ray energy is then
\begin{equation}
E_{K_{\alpha}}^\textrm{BS}=E_2-E_1=m_ec^2\left( \sqrt{1-\alpha^2(Z-1)^2/4} - \sqrt{1-\alpha^2(Z-1)^2} \right) \label{eq:BS}.
\end{equation}
 Bohr-Sommerfeld calculations are not exact --- they are approximations to more accurate wave mechanics calculations. Remarkably, however, they do match the result of the exact wave mechanical calculation in the case of circular orbits. In the nonrelativistic case, the correct wave equation to use would be the Schrodinger's equation. In the relativistic case, the correct one would be Dirac's equation. However, in both cases, the Bohr-Sommerfeld approximations are much more accessible for student understanding than solutions to the full wave equations. 

\section{Apparatus and procedure}

\begin{figure}
\includegraphics[width=9cm]{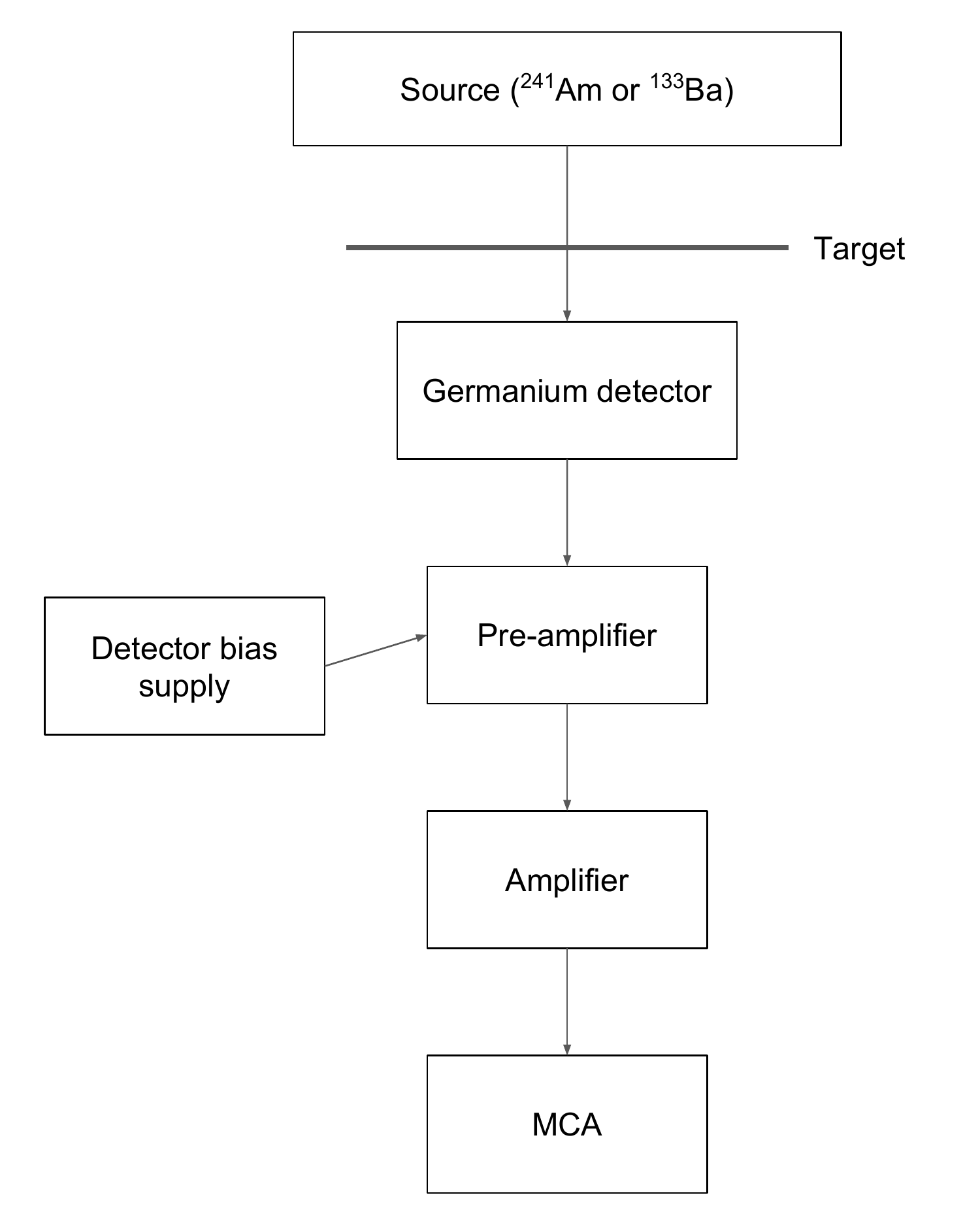}
\caption{Block diagram of the apparatus and signal chain. The detector contains ap-n doped, reverse-biased, nitrogen-cooled germanium crystal that generates pulses of electrical charge proportional to the energy deposited by each incident x ray. These pulses are integrated to a voltage by the pre-amplifier, amplified, and counted in amplitude bins by the MCA.}
\label{fig:block-diagram}
\end{figure}

To test Moseley's law, we measure the energy of $K_{\alpha}$ radiation for a variety of elements. To do this, we expose selected elemental samples to radiation from high-energy sources, inducing the emission of characteristic x rays. The x rays are then measured by a high resolution energy detector, generating a counting signal that is recorded as a spectral histogram by a multichannel analyzer (MCA), as in Fig.~\ref{fig:block-diagram}. Once the system's energy sensitivity is calibrated, we use the MCA's output to determine the energy of the x rays which hit the detector by identifying MCA histogram channels with distinct peaks in the counting rate.

\subsection{Detector and calibration}

Measurements are performed a Canberra model BE2020 broad energy solid state x-ray detector system.\cite{detector} The detector itself is a crystal of p-n doped germanium, biased at $-1300$~volts. The crystal is cooled with liquid nitrogen at 77~K to improve energy resolution by reducing thermal noise. At the energy range we are studying, x-ray photons excite thousands of electrons in the germanium through photoelectric absorption. These electrons, along with a few electrons excited through thermal noise, enter the conduction band of the germanium crystal, producing a current signal in the detector. Since the number of excited electrons is proportional to the energy of the incident x ray, the magnitude of the integrated current signal is also proportional to the energy of the x ray. 

As shown in Fig.~\ref{fig:block-diagram}, the current signal from the germanium detector passes through a charge-integrating pre-amplifier and voltage amplifier to convert it to a voltage signal in the range 0--10~V, which can then be recorded by an MCA. The MCA bins the signal into one of 2048 channels based on the voltage amplitude of the signal. The corresponding x-ray energy of each MCA channel should then vary linearly with the channel number. This relationship between energy and channel number can be calibrated by exposing the detector system to sources of high-energy photons with well-known and well-defined energies.

\begin{figure}
\includegraphics[width=9cm]{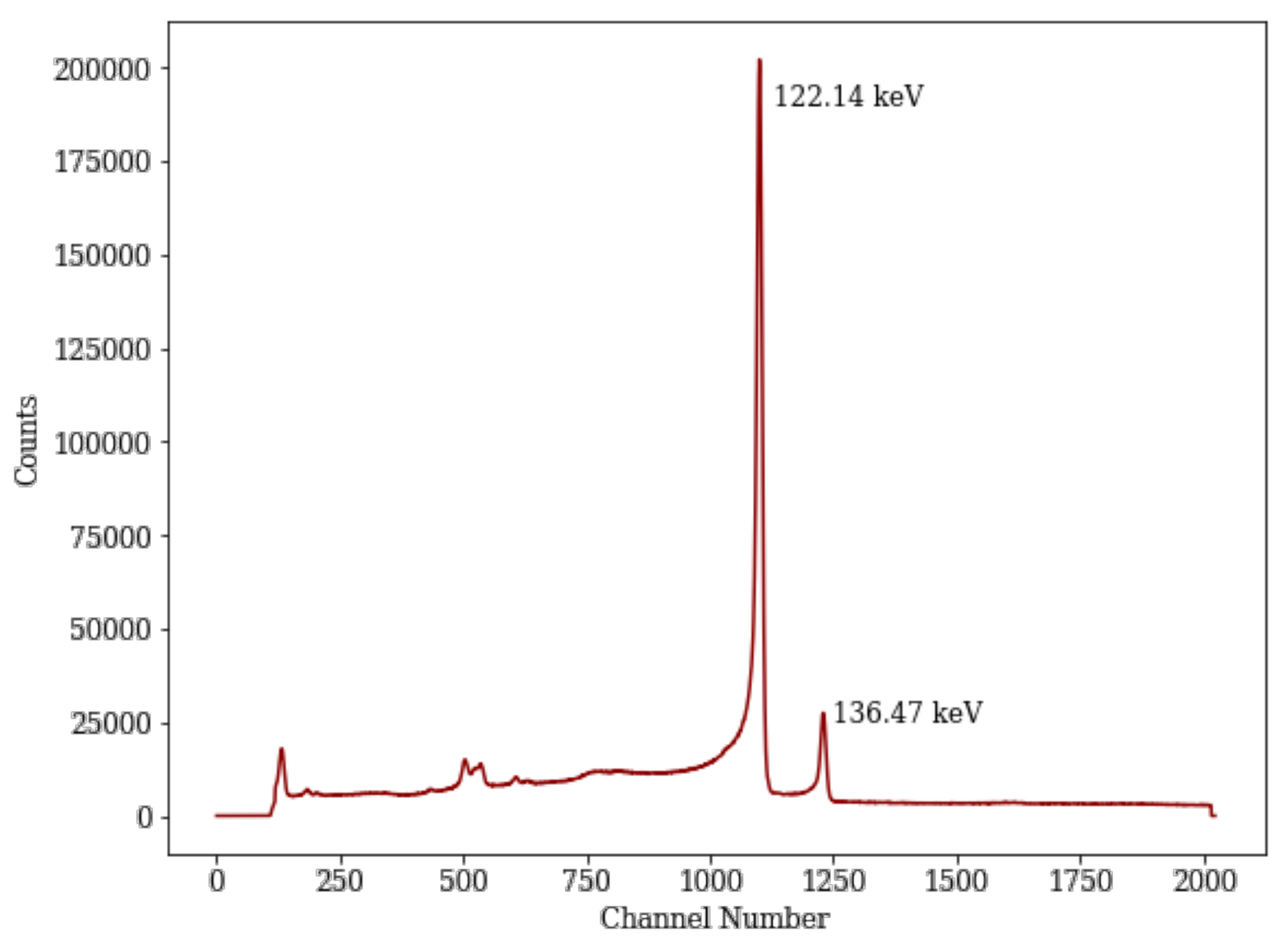}
\caption{An example MCA calibration spectrum of an isotope ($^{57}$Co in this case) with several well-known gamma ray emission energies. The data shown is a histogram of x-ray counts versus MCA channel number after five minutes of sampling.} 
\label{fig:cobalt}
\end{figure}

We calibrated the MCA using several small radioactive sources: $^{57}$Co, $^{133}$Ba, and $^{137}$Cs, each with several micro-Curies of activity; see Fig.~\ref{fig:cobalt}. These isotopes were chosen because they emitted sharp gamma radiation peaks across the range from near 10~keV to just over 100~keV, the same range as the x-ray energies to be measured. We collected data on these calibration spectra until we saw clean peaks, typically after a few minutes. We then calibrated the MCA channels to those peaks via a linear fit using $\chi^2$ minimization. The exact energy per channel depended on amplifier gain and detector bias settings that could vary between experimental runs, so calibration was repeated each time. Typical values ranged between 0.05--0.10~keV per channel.

\subsection{X ray sources and targets}

\begin{figure}
	\begin{subfigure}[b]{0.5\textwidth}
		\includegraphics[width=0.75\textwidth]{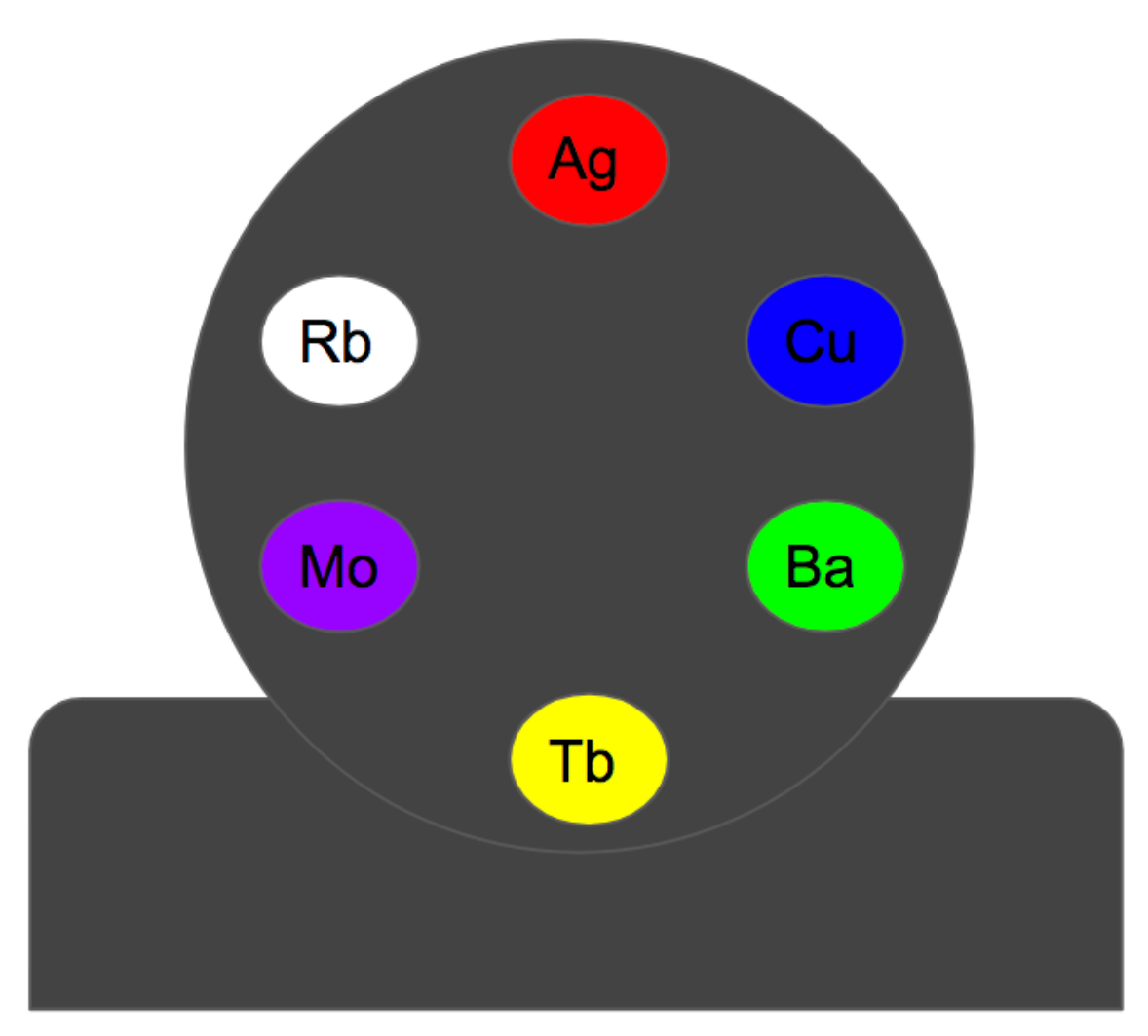}
		\caption{Front view of the Amersham AMC.2084 x-ray source.}
    	\label{fig:am-front}
	\end{subfigure}%
	\begin{subfigure}[b]{0.5\textwidth}
		\includegraphics[width=\textwidth]{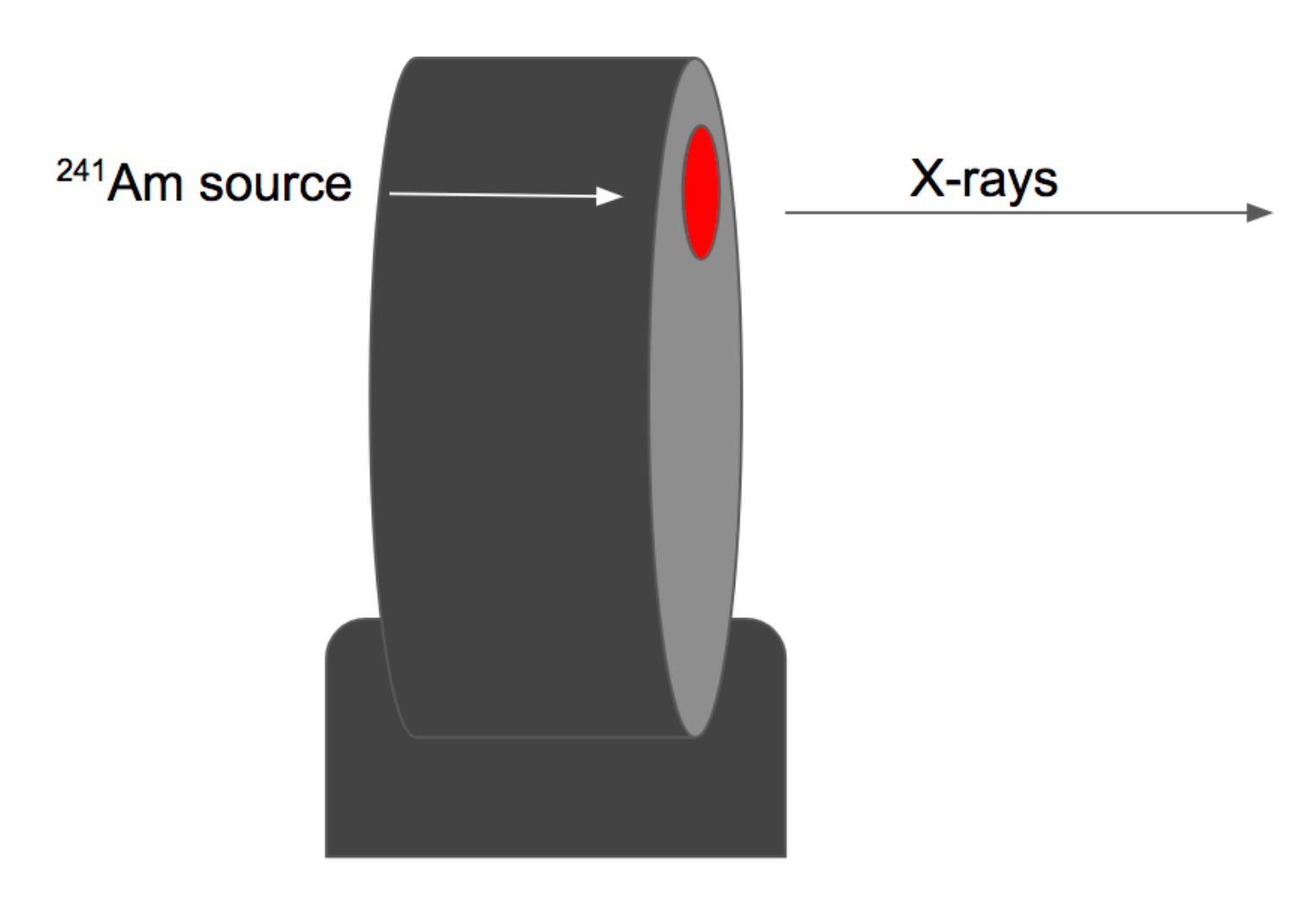}
		\caption{Side view of the Amersham AMC.2084 x-ray source.}
		\label{fig:am-side}
	\end{subfigure}
	\caption{The Amersham AMC.2084 10~mCi $^{241}$Am variable energy x-ray source\cite{amersham} contains six different metals on a rotatable wheel, each of which has its own characteristic x-ray spectrum. Radiation from the source, in the form of 5.48~MeV alpha particles and 59.5~keV gamma rays, bombards the metal samples, causing them to fluoresce x rays which exit the assembly towards a detector.}\label{fig:amersham}
\end{figure}

We used two high-energy radiation sources, $^{241}$Am (10~mCi) and $^{133}$Ba (7~$\mu$Ci), to generate x rays from pure samples of various elements. The $^{241}$Am source is part of a variable energy x-ray source assembly.\cite{amersham} It emits alpha particles with energies near 5.48~MeV and gamma rays near 59.5~keV as it decays to $^{237}$Np. The alpha and gamma radiation bombard one of six metals in a rotatable wheel, causing the metals to fluoresce characteristic x rays which exit the assembly as a beam (see Fig.~\ref{fig:amersham}). The experimenter can rotate the wheel to select different metals, generating x rays from copper, rubidium, molybdenum, silver, barium, and terbium. 

In addition to this variable x-ray source, we also used a $^{133}$Ba source that emits x rays at around 80~keV (as well as gamma rays with a few hundred keV each)\cite{radioactiveBa} to bombard prepared metal targets of high purity. We used this technique to measure tantalum, tungsten, platinum, gold, and lead --- all of which have x-ray lines below 80 keV. In this setup, the gold and platinum targets are foils held together by kapton tape. (We also bombarded a pure ball of kapton tape with radiation from the $^{133}$Ba source in order to ensure that the tape did not affect the measured peaks.)

The final sample was uranium. The uranium sample came not from a pure elemental sample as above, but rather from a red-orange glazed Fiesta brand ceramic dinnerware plate. The bright red-orange glaze (branded ``Fiesta red'') on these ceramics produced in the years 1936--1943 contains natural uranium oxide, while those produced in the years 1959--1972 contain the oxide of isotopically depleted uranium.\cite{fiesta1,fiesta2} Whether the present sample is of depleted or natural uranium has not been determined. Regardless, the sample's own radioactivity is enough to induce x-ray fluorescence. 

The sample was left in the detector for three days in order to determine the peak to within one channel. This is an important data point because of its high atomic number ($Z$=92), which gives a larger deviation from nonrelativistic theories. It is worth noting that there are several additional sources of error in the case of the uranium sample. These will be discussed in the next section. 

Note the wide range of atomic numbers and the substantial gap between the second heaviest element, lead, and the heaviest element, uranium. The relativistic model predicts that the innermost electrons in copper are moving with velocity $0.21c$ ($\gamma=1.02$), while the innermost electrons in uranium are moving with velocity $0.67c$ ($\gamma=1.35$), well into the relativistic regime.

\section{Data and analysis}
Once the system was calibrated, we used it to measure x rays produced by the methods described above. Table~\ref{table} shows the data collected for each element. Each point has a statistical error of about 0.06~keV due to the resolution of the peaks on the MCA. 

\begin{table}
\begin{center}
\caption{Measured $K_\alpha$ x-ray energies for each element tested. Each peak energy has a statistical uncertainty of about 0.06~keV. }
 \begin{tabular}{c c c} 
\hline\hline
Atomic number & Element & Peak (keV)\\
\hline
29 & Cu & 8.1\\
37 & Rb & 13.45\\
42 & Mo & 17.46\\
47 & Ag & 22.14\\
56 & Ba & 32.2\\
65 & Tb & 44.4\\
73	 & Ta & 57.5\\
74 & W & 59.3\\
78 & Pt & 66.7\\
79 & Au & 68.6\\
82 & Pb & 74.9\\
92 & U & 98.26\\
\hline\hline
\end{tabular}
\label{table}
\end{center}
\end{table}

In addition to statistical errors mentioned above, the uranium plate introduces several sources of possible error. First, the plate contains an oxide rather than elemental uranium. Since the experiment is mostly concerned with inner shell phenomena, we need to know if the valence-shell bonding would matter. Since the valence electrons are several times further away then the nucleus and have far less effective charge, we estimate that they cannot induce an error of more than one percent.

\begin{figure}
\centering
\begin{subfigure}[b]{0.5\textwidth}
\includegraphics[width=\linewidth]{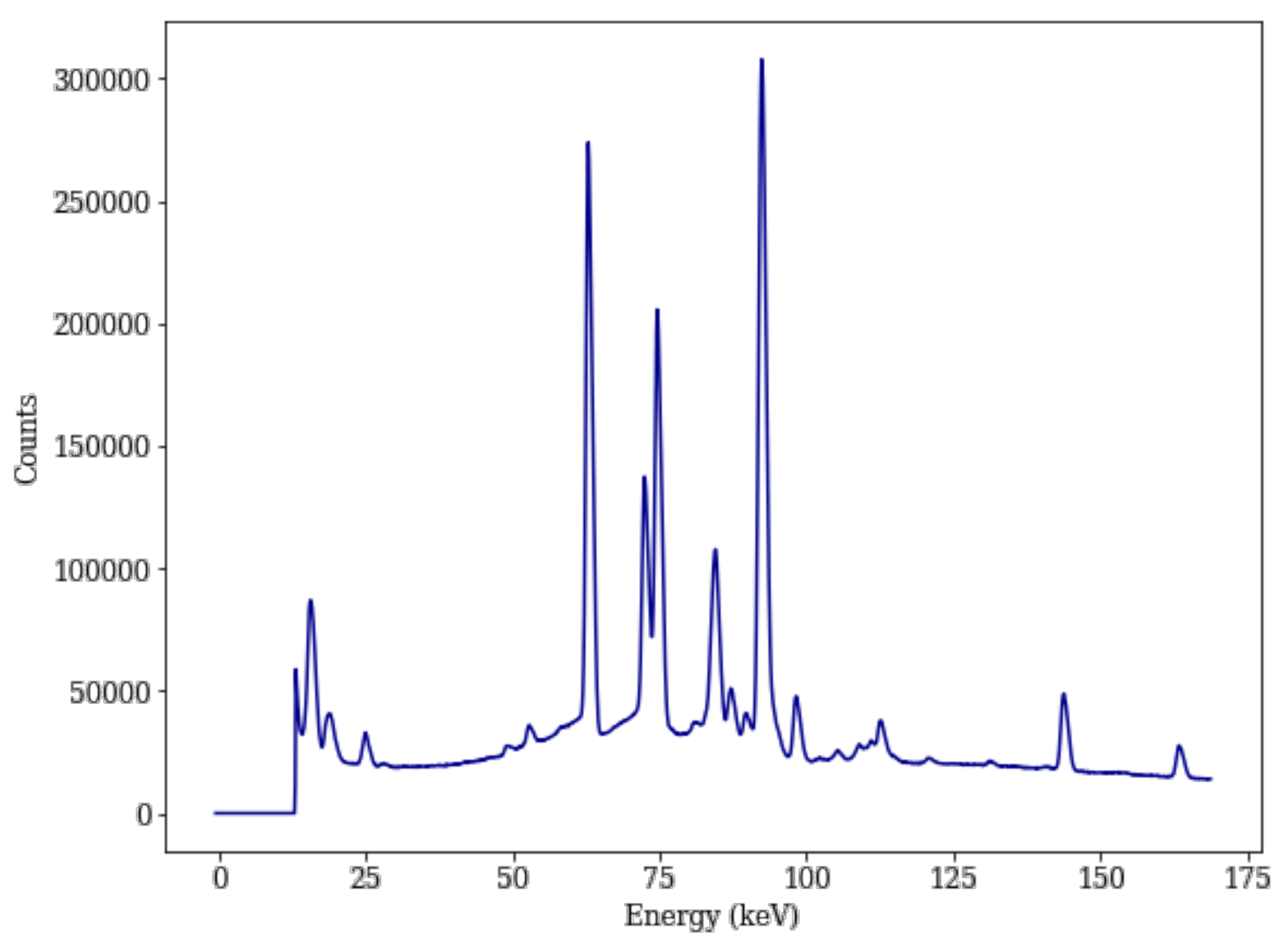}
\caption{The many peaks shown on the MCA from the uranium-containing Fiesta ceramic plate.}
\label{fig:uf}
\end{subfigure}%
\begin{subfigure}[b]{0.5\textwidth}
\includegraphics[width=\linewidth]{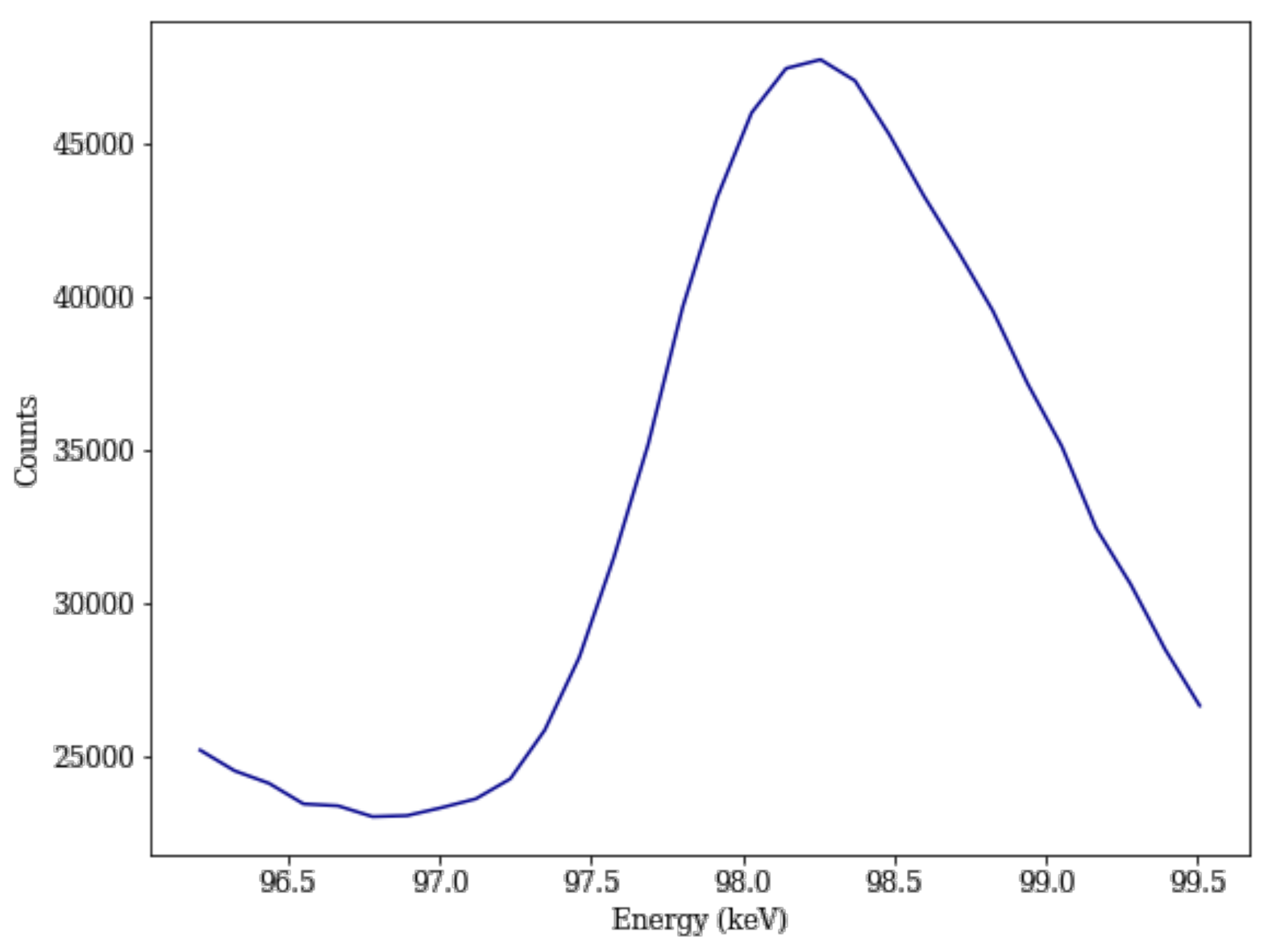}
\caption{A close up of the peak used to determine the uranium $K_\alpha$ energy.}
\label{fig:uc}
\end{subfigure}
\caption{}
\end{figure}

Another error may be introduced by the spectrum's many peaks, as seen in Fig.~\ref{fig:uf}. The plate contains uranium, a host of decay products, and many other fluorescing materials. It is possible that there was another peak overlapping with that of the uranium. However, the peak, shown in Fig.~\ref{fig:uc}, is narrow, with a full width at half maximum of about 2.3~keV. The global maximum of counts falls between two adjacent channels, but the counts for these channels are higher than the counts for either of the next lower or higher energy channels with Poisson-based probability of $1.3\times10^{-12}$ and $2.4\times10^{-4}$ respectively. Because of this, we believe the error introduced by possible overlap is at most one channel. This lineshape-model-agnostic analysis takes advantage of the high number of counts in the integration to avoid additional errors which could be introduced by a theory bias.

\begin{figure}
\centering
\begin{subfigure}[b]{0.5\textwidth}
\includegraphics[width=\linewidth]{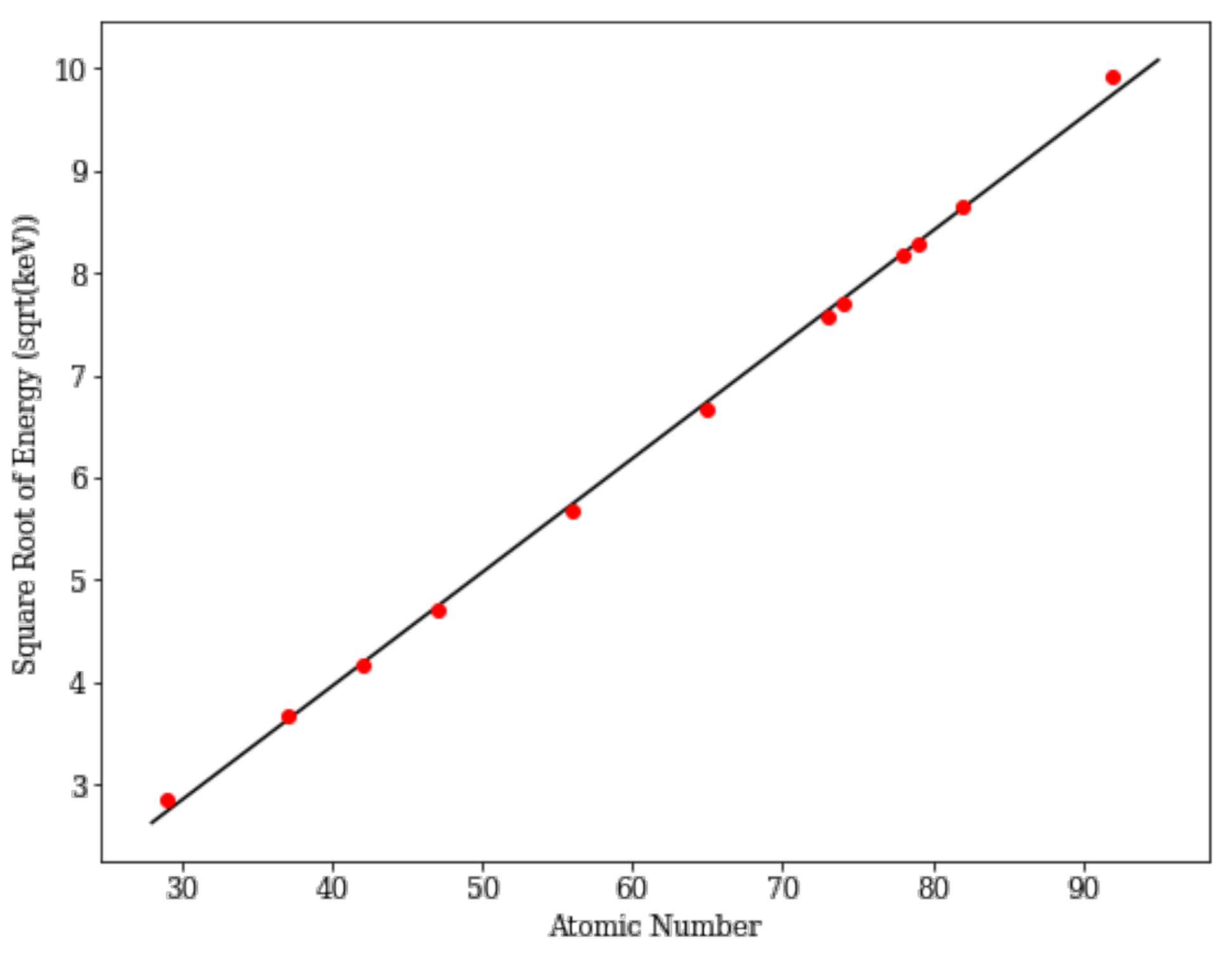}
\caption{Linear fit suggested by Moseley's law.}
\label{fig:ZrootE}
\end{subfigure}%
\begin{subfigure}[b]{0.5\textwidth}
\includegraphics[width=\linewidth]{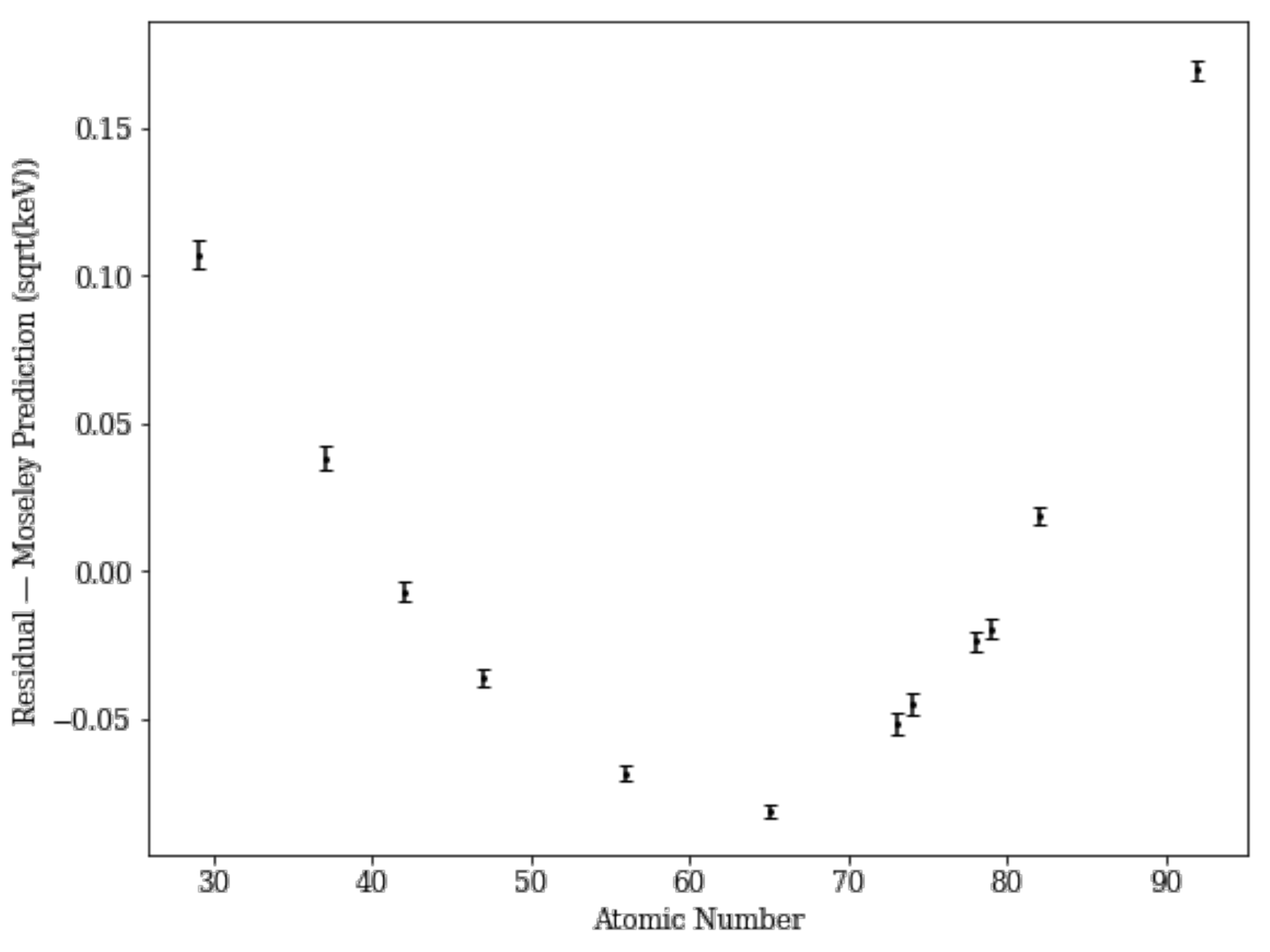}
\caption{Residuals of data to the Moseley's law fit. }
\label{fig:MosRes}
\end{subfigure}
\caption{The visually apparent good fit in Fig.~\ref{fig:ZrootE} is shown to be spurious by the obviously nonrandom trend in the residuals of Fig.~\ref{fig:MosRes}.} 
\end{figure}

Fig.~\ref{fig:ZrootE} shows a linear fit  between the atomic number and the square root of the energy, testing Moseley's original law and the Bohr model of Eq.~(\ref{eq:bohr}). Our fit suggests an electron mass of $593\pm 10_{\textrm{stat}}\pm 6_{\textrm{sys}}$~keV/$c^2$, which is more than seven standard deviations away from values found in other experiments.\cite{pdb} Visually, the line appears to be a good fit, but the $\chi^2$ value is over 5880 with just 10 degrees of freedom --- corresponding to a probability of less than $10^{-1000}$, which we consider negligible --- suggesting a bad fit and providing a valuable learning opportunity.

To verify that the theory behind Moseley's law is inaccurate, we plot the residuals to this linear fit in Fig.~\ref{fig:MosRes}. The clear pattern in the residual supports our belief that a new model is required.
 
\begin{figure}
\includegraphics[width=9cm]{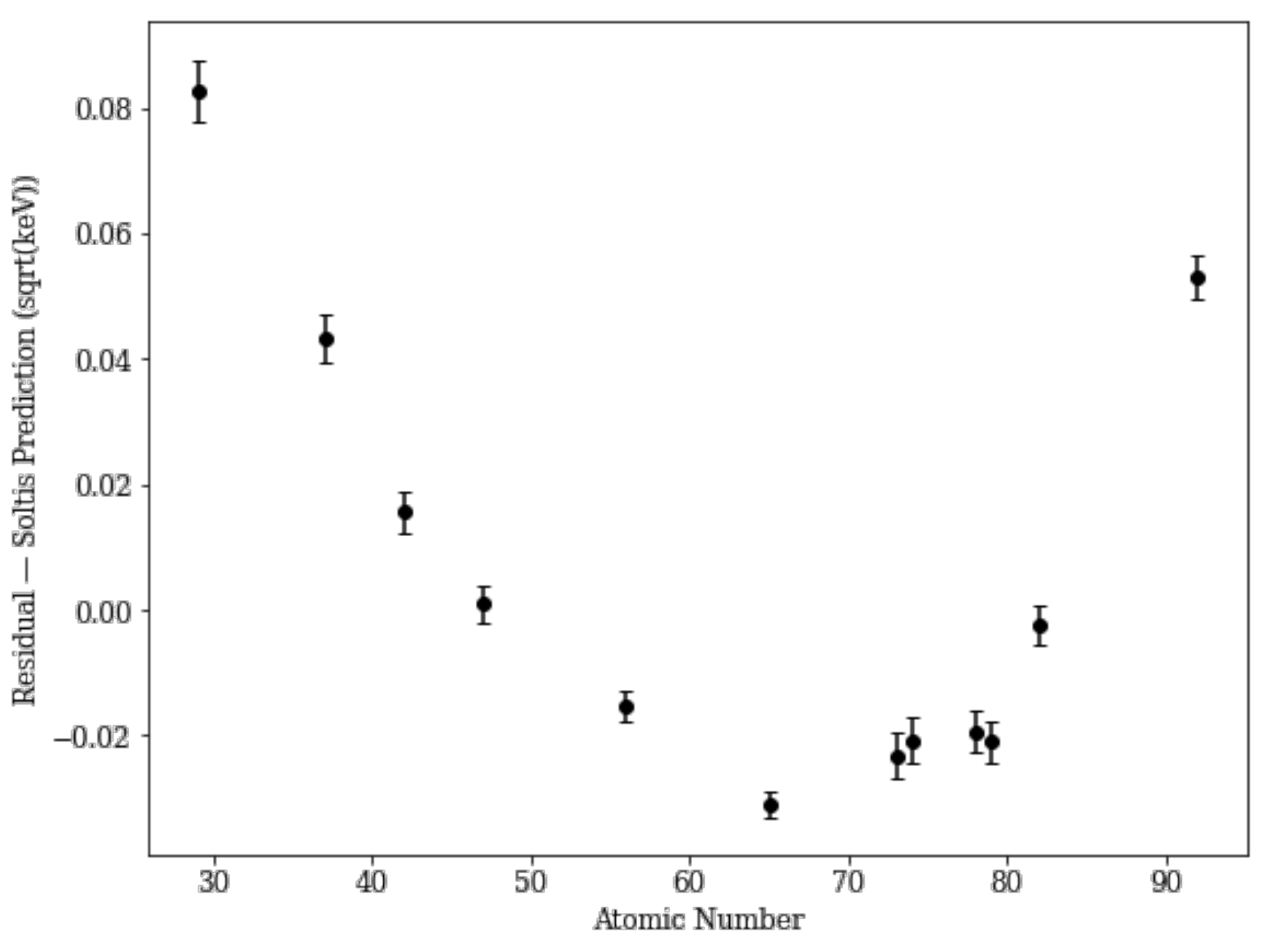}
\caption{Residuals of data to fit suggested by Soltis \textit{et al}.\cite{soltis2017} An obvious nonlinear trend is apparent, suggesting the model does not represent the data well.}
\label{fig:soltRes}
\end{figure}

We next test the model described by Soltis \textit{et al.}\cite{soltis2017}, that is, the first-order perturbative corrections to the semiclassical Bohr model, as in Eq.~(\ref{eq:soltis}). We again see a strong trend in the residuals, shown in Fig.~\ref{fig:soltRes}, and find a high $\chi^2$ value: 1077 with 10 degrees of freedom, corresponding to a probability value of about $10^{-200}$. While this is clearly an improved fit over Moseley's law, the large $\chi^2$ value provides motivation to search for a still better model.

\begin{figure}
\includegraphics[width=9cm]{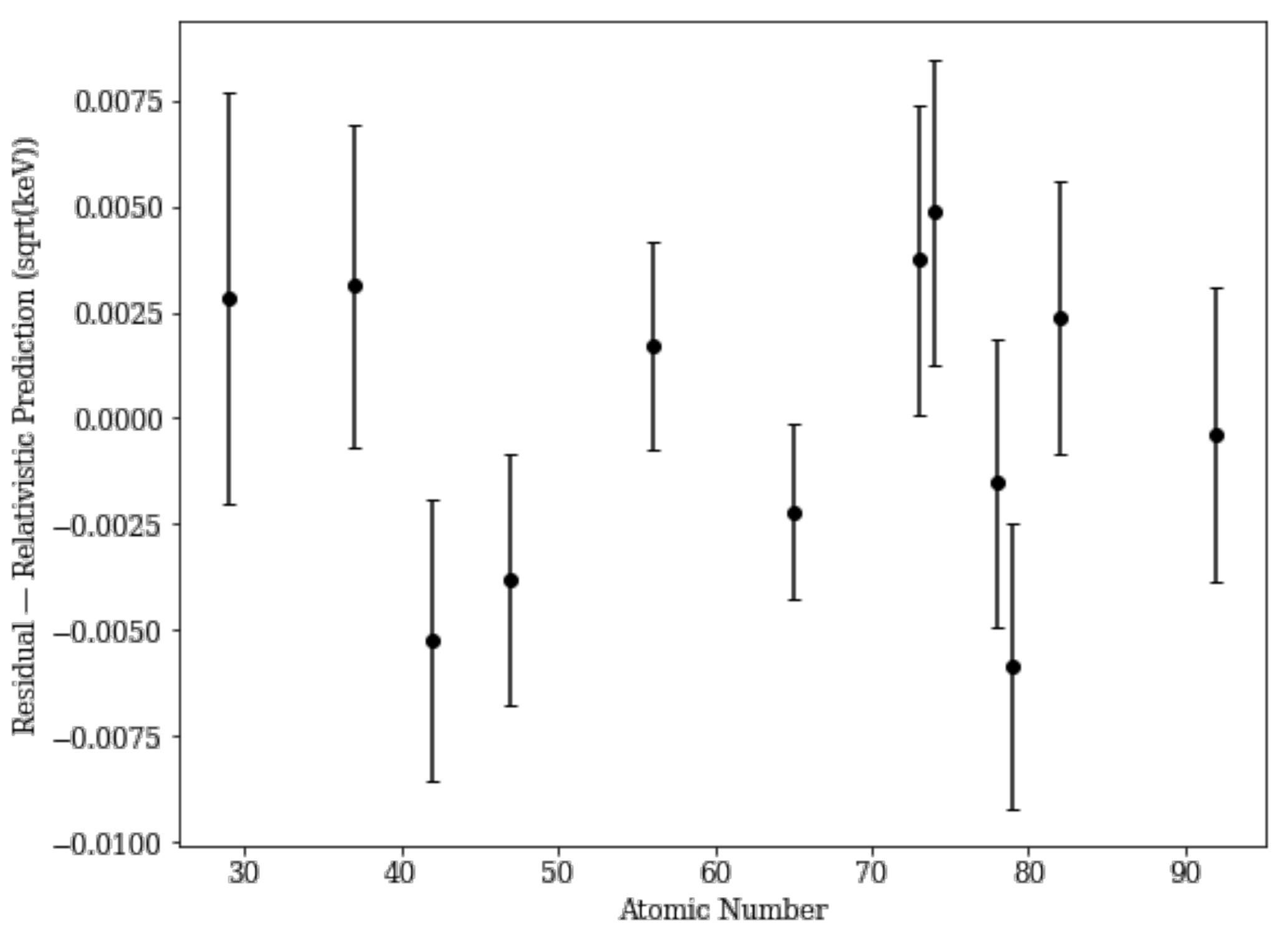}
\caption{Residuals of data to relativistic Bohr-Sommerfeld approximation fit. No obvious trend is apparent.}
\label{fig:RelRes}
\end{figure}

We finally use the fit presented in Eq.~\ref{eq:BS} to find $m_ec^2$ and test our claim that the fully relativistic Bohr-Sommerfeld approximation provides a more accurate model. The residuals from this fit are shown in Fig.~\ref{fig:RelRes}, which indicates no significant trend. This fit has a $\chi^2$ value of 13.4 and 10 degrees of freedom, consistent with random noise, indicating that the fully relativistic theory is the most accurate of the models tested. The probability value of the relativistic hypothesis is 0.20, so it is not rejected.  In addition, we find that $m_ec^2=502 \pm 6$~keV, only 1.5 standard deviations from the well-known value of $511$~keV found in other experiments.\cite{pdb}

\section{Conclusions}
This paper examines a correction to Moseley's law that accounts for relativistic effects. It shows that the corrected version better explains measured data than both the nonrelativistic version and alternative theories tested in the past. The nonrelativistic version of the law gives an electron mass inconsistent with the other experiments while the relativistic version agrees within 1.5 standard deviations.

All of the samples measured in this experiment are metals. It would be interesting to include some heavy elements which are not metals --- for example, an iodine tablet. This would allow us to determine whether different categories of elements had different x-ray behavior.

This experiment is a valuable teaching opportunity, as it requires experimenters to look at residual plots to clearly reveal incorrect models.  In addition, the theory presented here is fundamentally both relativistic and quantum. That means that this experiment not only demonstrates that a well-known model, Moseley's law, is inaccurate, but it also tests both relativity and quantum mechanics at the same time. The combined theory is accurate to the limit of the experimental apparatus and allows students to explore the process of discovering new explanations as data becomes more accurate.

\begin{acknowledgments} This experiment was collaborative across both space and time, and we would like to thank all involved both in making the original observation and generating the final results written up here. We would especially like to thank the MIT Junior Lab staff for lending their time, expertise, and dinnerware to the experiment. 
\end{acknowledgments}


%



\begin{thebibliography}{11}%

\bibitem{moseley1}%
   H.G.J.~Moseley, 
   ``\href{https://www.tandfonline.com/doi/abs/10.1080/14786441308635052}{The high-frequency spectra of the elements},''
   Philos.~Mag.~Series~6 \textbf{26}~(156), 1024--1034 (1913).

\bibitem{moseley2}%
   H.G.J.~Moseley, 
   ``\href{https://www.tandfonline.com/doi/abs/10.1080/14786440408635141}{The high-frequency spectra of the elements. Part II},''
   Philos.~Mag.~Series~6 \textbf{27}~(160), 703--713 (1914).

\bibitem{Bohr}%
   Niels~Bohr, 
   ``\href{https://www.tandfonline.com/doi/abs/10.1080/14786441308634955}{On the Constitution of Atoms and Molecules},''
   Philos.~Mag.~Series~6 \textbf{26}~(151), 1--25 (1913).

\bibitem{soltis2017}%
  Tomaz~Soltis, Lorcan~M.~Folan, and Waleed~Eltareb,
  ``\href{https://aapt.scitation.org/doi/10.1119/1.4977793}{One hundred years of Moseley's law: An undergraduate experiment with relativistic effects},''
  Am.~J.~Phys. \textbf{85}, 352--358 (2017).

\bibitem{pdb}%
   C.~Patrignani \textit{et al.}\ (Particle Data Group),
   ``\href {http://pdg.lbl.gov}{Review of Particle Physics},''
   Chin.~Phys.~C \textbf{40}~(10), 1--1808 (2016).

\bibitem{kraft1974}%
  David Kraft,
  ``\href {https://aapt.scitation.org/doi/10.1119/1.1987875}{Relativistic Corrections to the Bohr Model of the Atom},''
  Am.~J.~Phys. \textbf{42}~(10), 837--839 (1974).

\bibitem{detector}%
   Canberra Industries, Inc.,
   \textit{Germanium Detectors: User's Manual}
   (2003).

\bibitem{amersham}%
   The Radiochemical Centre (Amersham),
  \textit{Variable energy X-ray source: code AMC.2084}, Data sheet 11196
  (June 1975).

\bibitem{radioactiveBa}%
  V.P.~Chechev and N.K.~Kuzmenko, ``\href{http://www.nucleide.org/DDEP_WG/Nuclides/Ba-133_tables.pdf}{133-Ba},'' in \textit{Table de radionucleides}, Decay Data Evaluation Project \texttt{<}\url{http://www.lnhb.fr/ddep_wg/}\texttt{>} (2016).

\bibitem{fiesta1}%
  The Hall China Company, 
  ``Color history'',
  \texttt{<}\url{https://fiestafactorydirect.com/pages/color-history}\texttt{>}, 
  [Retrieved 2018-09-04].

\bibitem{fiesta2}%
  The Homer Laughlin China Co, 
  ``Radiation in ceramic glazes, 2011-03-16'',
  \texttt{<}\url{https://web.archive.org/web/20120401000958/http://www.hlchina.com/gmastatement.html}\texttt{>},
  Archived 2014-04-16 at the Wayback Machine,  [Retrieved 2018-09-40].
\end{thebibliography}
\end{document}